# Collapsed speech segment detection and suppression for WaveNet vocoder

*Yi-Chiao Wu[1], Kazuhiro Kobayashi[2], Tomoki Hayashi[3], Patrick Lumban Tobing[1], Tomoki Toda[2]*

[1] Graduate School of Informatics, Nagoya University, Japan
[2] Information Technology Center, Nagoya University, Japan
[3] Graduate School of Information Science, Nagoya University, Japan
`yichiao.wu@g.sp.m.is.nagoya-u.c.jp`, `tomoki@icts.nagoya-u.c.jp`

## Abstract

In this paper, we propose a technique to alleviate the quality degradation caused by collapsed speech segments sometimes generated by the WaveNet vocoder. The effectiveness of the WaveNet vocoder for generating natural speech from acoustic features has been proved in recent works. However, it sometimes generates very noisy speech with collapsed speech segments when only a limited amount of training data is available or significant acoustic mismatches exist between the training and testing data. Such a limitation on the corpus and limited ability of the model can easily occur in some speech generation applications, such as voice conversion and speech enhancement. To address this problem, we propose a technique to automatically detect collapsed speech segments. Moreover, to refine the detected segments, we also propose a waveform generation technique for WaveNet using a linear predictive coding constraint. Verification and subjective tests are conducted to investigate the effectiveness of the proposed techniques. The verification results indicate that the detection technique can detect most collapsed segments. The subjective evaluations of voice conversion demonstrate that the generation technique significantly improves the speech quality while maintaining the same speaker similarity.

## 1. Introduction

In recent years, several works have focused on speech synthesis using deep neural networks (DNNs) [1-8]. One of the state-of-the-art techniques is WaveNet [8], which is an autoregressive model for predicting the probability distribution of a current waveform sample based on a specific number of previous samples. Because of the data-driven characteristic of WaveNet, it directly generates raw audio samples without various assumptions based on prior knowledge specific to audio (e.g., source-filter modeling for normal speech). Moreover, WaveNet has been applied to many applications, such as speech enhancement [9, 10], text-to-speech (TTS) synthesis [3, 5], singing voice synthesis [11], and speech coding [12]. In this paper, we focus on the use of WaveNet as a vocoder [13-16] to replace conventional vocoders (e.g., WORLD [17, 18]), which are usually developed on the basis of the source-filter model and cause significant degradation in the naturalness of speech.

In [13-16], the effectiveness of the WaveNet vocoder for generating natural speech with acoustic spectral and prosodic features as auxiliary features was demonstrated. However, in [14, 19, 20], we found that the samples generated from the WaveNet vocoder sometimes become unstable. This instability causes the amplitudes of some generated speech segments to suddenly become extremely large and contain almost equal intensity at all frequencies, similarly to white noise. Consequently, the resulting speech sample has significantly degraded naturalness. One possible reason for this problem is the difference in the acoustic features used for training and decoding. Specifically, this problem has been observed in our proposed voice conversion (VC) system [14, 19, 20] with the WaveNet vocoder, which is a technique to convert the speaker identity of speech while maintaining the same linguistic contents. In the proposed VC system, the acoustic features of the source speaker are converted into those of the target speaker using a statistical mapping function separate from the WaveNet vocoder. Therefore, instead of directly generating a waveform based on the acoustic features of natural speech, the WaveNet vocoder generate samples with the converted acoustic features, and these less accurately predicted acoustic features easily cause the collapsed speech problem. Moreover, because of the limited source-target parallel corpus, it is still difficult to directly train the WaveNet vocoder using converted acoustic features.

A straightforward way to tackle the collapsed speech problem is to constrain the predicted probability distribution by using another probability distribution derived from linear predictive coding (LPC) coefficients [19, 20] to prevent WaveNet from generating extremely discontinuous or non-speech-like samples. However, this LPC constraint also introduces an over-smoothing effect into the generated speech because the LPC coefficients are estimated from over-smoothed converted acoustic features. Therefore, in [19, 20], we proposed a system selection framework with collapsed speech detection based on an utterance-based power difference. The framework only selected the utterances generated with the LPC constraint when the utterances generated from the original WaveNet vocoder suffered from collapsed speech.

In this paper, we propose an improved segment-based collapsed speech detector that is based on envelope detection [21] and can detect more collapsed types than the previous method. In this method, collapsed segments are first detected, and an LPC-based constraint for WaveNet vocoder is only applied to the collapsed segments. Therefore, we can restrict the over-smoothing effect only in a short time slot, which avoids the quality degradation caused by over-smoothing. Verification and subjective tests are conducted to evaluate the detection performance of the proposed detectors, and the speech quality and similarity of the proposed system, respectively.

## 2. WaveNet vocoder

WaveNet [8] is a deep autoregressive network capable of directly modeling a speech waveform sample-by-sample using the following conditional probability:

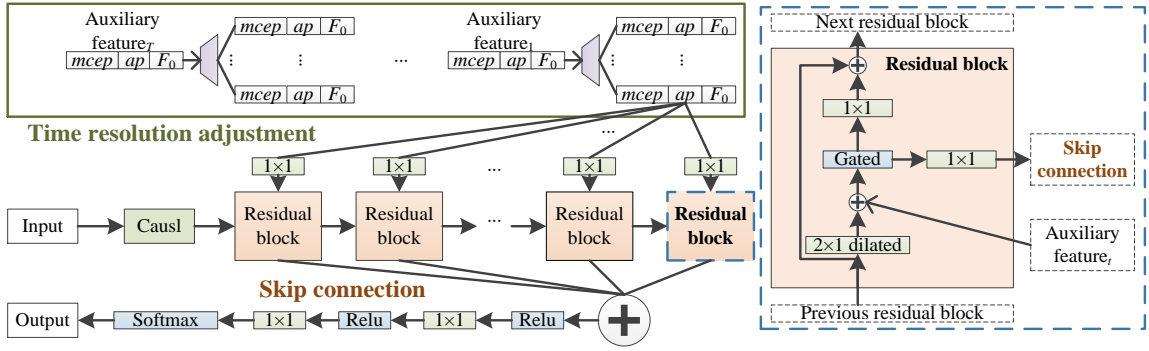

Figure 1: *Conditional WaveNet vocoder architecture*

$$P(\mathbf{Y} \mid \mathbf{h}) = \prod_{n=1}^{N} P(y_n \mid y_{n-r}, ..., y_{n-1}, \mathbf{h}), \quad (1)$$

where $n$ is the sample index, $r$ is the size of the receptive field (a specific number of previous samples), $y_n$ is the current audio sample, and $\mathbf{h}$ is the vector of the auxiliary features. In our system, $\mathbf{h}$ consists of the coded aperiodicity (*ap*), the transformed $F_0$, and the converted spectral features (*mcep*).

Figure 1 shows the structure of the WaveNet vocoder, which consists of many residual blocks, each containing a 2 × 1 dilated causal convolution, a gated activation function, and two 1 × 1 convolutions. The dilated causal convolution is a skipping value filter, which enables the network to efficiently operate on a large receptive field. The gated activation function is formulated as

$$\tanh\left(\mathbf{V}^{(1)}_{f,k} * \mathbf{Y} + \mathbf{V}^{(2)}_{f,k} * a(\mathbf{h})\right) \odot \sigma\left(\mathbf{V}^{(1)}_{g,k} * \mathbf{Y} + \mathbf{V}^{(2)}_{g,k} * a(\mathbf{h})\right), \quad (2)$$

where $\mathbf{V}^{(1)}$ and $\mathbf{V}^{(2)}$ are trainable convolution filters, $*$ is the convolution operator, $\odot$ is an elementwise multiplication operator, $\sigma$ is a sigmoid function, $k$ is the layer index, $f$ and $g$ represent the "filter" and "gate", respectively, and $a(\cdot)$ is the resolution adjustment function used to duplicate auxiliary features to match the resolution of input speech samples. The input waveform is quantized to 8 bits on the basis of $\mu$-law encoding and the generated waveform is restored by $\mu$-law decoding.

## 3. Segment-based collapsed speech detection and suppression

For collapsed speech detection, the main concept of the proposed method is that even without listening to audio samples, people still can easily detect collapsed speech segments from the waveform shape. Furthermore, in accordance with the observation in our previous work [19], the quality of WaveNet-vocoder-generated speech is usually higher than that of speech generated from the WORLD vocoder, but the WaveNet vocoder is more sensitive to less accurately converted acoustic features. Moreover, although the perceptual qualities are different, the waveform envelopes and powers of the utterances generated by these vocoders are similar except for the collapsed speech segments. As a result, for the same acoustic features, it is reasonable to take the utterance generated from the WORLD vocoder as the reference to evaluate whether or not the utterance generated from the WaveNet vocoder contains collapsed speech segments. Moreover, owing to the similar waveform characteristics, the LPC coefficients extracted from the WORLD-generated utterances are also used to design a constraint for the WaveNet vocoder to avoid generating collapsed speech segments.

Figure 2 shows the procedure of the proposed system. After obtaining the acoustic features converted from the VC model, the proposed system uses the WORLD vocoder to synthesize the reference speech. Then, the reference envelope and LPC coefficients are extracted from the reference speech. As shown in Figure 2, every time the WaveNet vocoder generates audio samples with a predefined length, the system extracts the envelope of the non-overlapping speech segment and compares it with the corresponding reference envelope segment to check whether the segment contains collapsed speech. If collapsed speech is detected, the system regenerates the collapsed segment with the LPC-constrained WaveNet vocoder.

### 3.1. Speech waveform extraction

Because we must extract the envelope during the waveform generation procedure, a method with a low computational cost is required. In [21], Jarne proposed a heuristic approach to obtain a signal envelope that contains three simple steps. As shown in Figure 3, the first step is to take the absolute value of the signal. In the second step, the absolute signal is divided into several non-overlapping slots with a predefined window length, and then peak detection is performed by replacing all the values in each signal slot with the maximum value of that signal slot. Finally, the peak detection results are processed with a low-pass filter. In our system, we modify the framework by replacing the taking of the absolute value with Hilbert transform, because the experimental results, which will be presented in Section 4, show better collapsed speech detection performance with this modification. Furthermore, after obtaining the envelopes of the WaveNet- and WROLD-generated speech segments, if the difference between these envelopes exceeds an empirical threshold, collapsed speech is detected.

### 3.2. LPC-constrained WaveNet vocoder

The main concept of LPC is that the current signal sample can be a linear combination of previous signal samples. That is, the LPC coefficients describe the relationship between the current sample and past samples, and we can apply the relationship to the probability distribution predicted from the WaveNet vocoder to avoid generating extremely non-speech-like samples. Specifically, in the LPC-constrained WaveNet vocoder, the equation used to constrain the predicted probability distribution of the current speech sample is

$$P(y_n | y_{n-r},...,y_{n-1},\mathbf{h},\phi) \propto$$
$$P(y_n | y_{n-r},...,y_{n-1},\mathbf{h})\left(P(y_n | y_{n-I},...,y_{n-1},\phi)\right)^\rho, \quad (3)$$

where $P(y_n | y_{n-r},...,y_{n-1},\mathbf{h})$ is the conditional probability from WaveNet given the vector of auxiliary features $\mathbf{h}$, $\rho$ is a control factor, and $P(y_n | y_{n-I},...,y_{n-1},\phi)$ is the probability mask from the LPC constraint, which is a probability mass function approximating a Gaussian distribution with mean $\mu_{lpc}$ and variance $\sigma_{lpc}$. The mean $\mu_{lpc}$ is the product of the past $I$ samples generated from the WaveNet vocoder with the corresponding LPC coefficients $\phi$ extracted from the reference samples generated from the WORLD vocoder. The variance $\sigma_{lpc}$ is the variance of the LPC prediction errors, which are extracted from the corresponding frame of the WORLD-generated sample. Furthermore, to simulate the distortion from $\mu$-law coding, the WORLD-generated samples are processed by $\mu$-law encoding and decoding before extracting the LPC coefficient. Note that we may also directly calculate the Gaussian distribution from the mel-cepstrum (auxiliary features for the WaveNet vocoder) by using the MLSA filter without generating the reference speech using the WORLD vocoder.

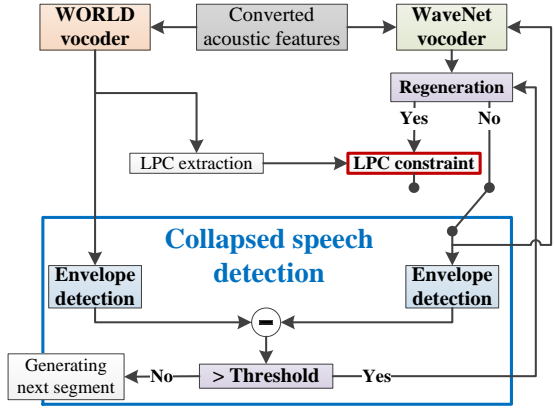

Figure 2: *Flowchart of the proposed system*

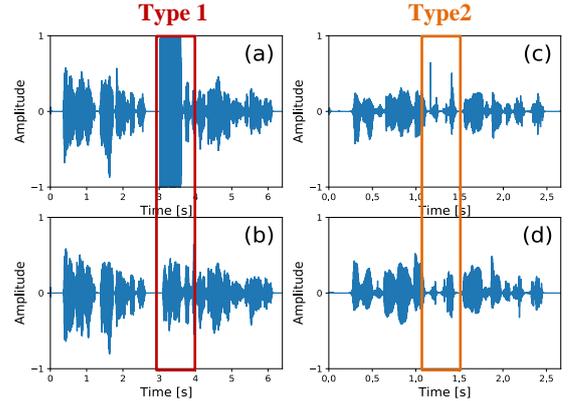

Figure 4: *(a) WaveNet-generated samples with **type-I** collapsed speech and (b) reference WORLD samples. (c) WaveNet-generated samples with **type-II** collapsed speech and (d) reference WORLD samples.*

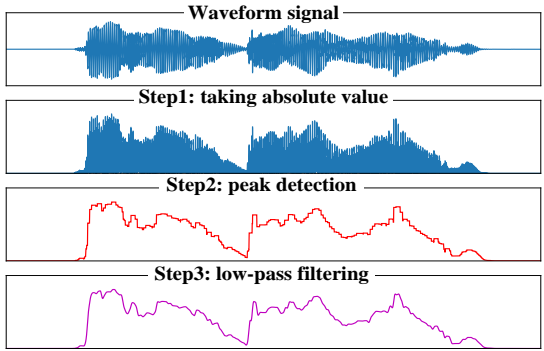

Figure 3: *Three steps of envelope detection*

## 4. Experiments

### 4.1. Experimental settings

We conducted objective and subjective tests on the SPOKE task corpus of Voice Conversion Challenge 2018 (VCC2018) [22], which was an English speech corpus. The SPOKE corpus included four source speakers and another four target speakers. Each speaker had 81 training utterances and 35 testing utterances, and the contents of the source and target utterances were non-parallel. Therefore, the total number of source-target pairs in the SPOKE task was 16, which included four female to female (F-F) pairs, four female to male (F-M) pairs, four male to female (M-F) pairs, and four male to male (M-M) pairs. The sampling rate of the speech signals was set to 22050 Hz and the quantization bit number was 16 bits.

We evaluated the performance of the proposed framework combined with our previous non-parallel VC system [19], which was a DNN-based two-stage VC system, and we generated the waveform using the WaveNet vocoder. Moreover, the hyperparameters and training procedures of the DNN and WaveNet models and the signal analysis settings also followed those in the previous work. Specifically, a noise-shaping technique [16] was also applied to the WaveNet vocoder. During the waveform generation stage, the LPC control factor $\rho$ was initially set as 0.01, and if the regenerated speech segment still contained collapsed speech, the system increased $\rho$ to 0.1 and 1. The length of the collapsed speech detection segment was set as 4000 samples, the length of the peak detection window was set as 200 samples, and the cutoff frequency of the low-pass filter in the envelope detection was set as 300 Hz, all these setting were decided empirically.

### 4.2. Evaluation of collapsed speech detection

We considered collapsed speech detection as a verification problem, so the performance of the detector was measured by false accept (collapsed utterances not being detected) rates and false reject (normal utterances being detected) rates. In addition, as shown in Figure 4, there were two types of collapsed speech, one had an extremely large power at all frequencies similarly to white noise (**type-I**), and the other one had irregular impulses over a very short time (**type-II**). Therefore, we evaluated the performances of detectors for both types on the basis of the following criterions:

- **ENV_/wNS_/HT**: The difference in the envelopes of the WaveNet- and WORLD-generated segments before/after noise shaping (NS) with/without Hilbert transform (HT).
- **maxMCD**: The maximum difference in the mel-cepstral distortions of the WaveNet- and WORLD-generated utterances to the converted mel-cepstral features.

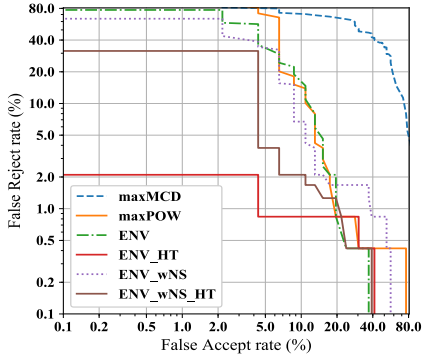

Figure 5: *DET curve for **type-I** detection*

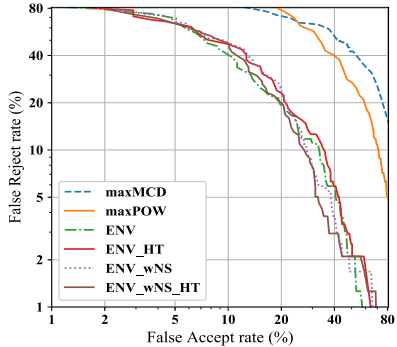

Figure 6: *DET curve for **type-II** detection*

- **maxPOW**: The difference in the maximum powers of the WaveNet- and WORLD-generated utterances.

For the evaluation dataset, a human subject labeled all the converted utterances generated by the WaveNet vocoder. The total number of converted utterances was 560, and 46 and 276 converted utterances were labeled as **type-I** and **type-II** collapsed utterances, respectively.

Figures 5 and 6 show the detection error tradeoff (DET) curves of these detectors. We found that the proposed methods (**ENV***) outperform **maxPOW** and **maxMCD** for both types of collapsed speech, particularly for the **type-II** speech. These results indicate that **maxPOW** can only detect the **type-I** collapsed speech, but the proposed method can detect both types. Moreover, replacing the absolute value with Hilbert transform improves the verification performance for **type-I** and achieves the similar performance for **type-II**. Additionally, the verification performances are similar for envelope detection before and after noise shaping. Therefore, we used the criterion based on envelope detection with Hilbert transform to detect the envelope of speech samples processed by noise shaping.

**4.3. Subjective evaluation**

The goal of the proposed framework is to generate samples without collapsed speech segments while maintaining the same speaker similarity as the original WaveNet vocoder. Therefore, we conducted a preference test to compare the quality of the waveforms generated from the WaveNet vocoder with and without the proposed framework. We also conducted similarity tests to evaluate the speaker identity conversion accuracy, in which listeners were given converted and target utterances and asked to select an answer from "definitely the same speaker", "probably the same speaker", "probably a different speaker" and "definitely a different speaker". The final similarity scores were the sum of the percentages of "definitely the same speaker" and "probably the same speaker".

Table 1: *Subjective results and p-values of the WaveNet vocoder with the proposed method (w/ CL) and without the proposed method (w/o CL)*

|  | *w/ CL* | *w/o CL* | *p*-value |
|---|---|---|---|
| **speech quality** | 77% | 23% | 1.091e-7 |
| **speaker similarity** | 46% | 48% | 0.813 |
| definitely the same | 11% | 10% | 0.810 |
| probably the same | 35% | 38% | 0.667 |
| probably different | 34% | 33% | 0.752 |
| definitely different | 20% | 19% | 0.963 |

For the evaluation data, we simultaneously generated waveforms with and without the use of the proposed collapsed speech detection and reduction framework, so when a collapsed speech segment was detected, the system outputted regenerated and non-regenerated speech samples. The total number of converted utterances was 560, 377 of which were detected containing collapsed segments based on the threshold chosen from the equal-error-rate point in Figure 6. However, because of the equal error rate of about 20% and the unoptimized threshold, the detection rate was much higher than that obtaining by human labeling. In the future, we can improve the detection performance by tuning the threshold. We randomly selected five utterance pairs for each speaker pair from the utterances in which collapsed segments was detected as the evaluation data in the quality test, and we randomly selected two utterance pairs for each speaker pair from the five selected pairs of utterances as the testing data in the similarity test. Therefore, the quality test contained 80 regenerated and 80 non-regenerated utterances, and the similarity test included 32 regenerated, 32 non-regenerated, and 32 natural target speech utterances. The number of subjective listeners was nine.

As shown in Table 1, the subjective results indicate that the proposed regeneration framework achieves a significant improvement in speech quality while maintaining similar speaker conversion accuracy. Specifically, the results prove that humans can be aware of collapsed speech segments and prefer regenerated samples without or with fewer collapsed speech segments. Furthermore, because we conducted the similarity test on converted and target utterances with the same contents, it made humans easily detect the similarity degradation caused by lower speech quality. Therefore, the similarity scores are much lower than the results of VCC2018, which were conducted on the utterances with different contents.

## 5. Conclusions

In this paper, we proposed a framework to detect collapsed speech segments and then regenerate the waveform with the LPC-constrained WaveNet vocoder to avoid the problem of collapsed speech. The proposed method achieved a high collapsed speech detection error rate and a 77 % preference in a subjective quality test while maintaining the same speaker similarity as that of the original WaveNet vocoder.

## 6. Acknowledgements

This work was partly supported by JST, PRESTO Grant Number JPMJPR1657, and JSPS KAKENHI Grant Number JP17H06101.